\documentclass{ieeeaccess}
\usepackage{cite}
\usepackage{amsmath,amssymb,amsfonts}
\usepackage{algorithmic}
\usepackage{graphicx}
\usepackage{textcomp}

\usepackage{subfigure}
\usepackage{booktabs}
\usepackage{framed}

\def\BibTeX{{\rm B\kern-.05em{\sc i\kern-.025em b}\kern-.08em
    T\kern-.1667em\lower.7ex\hbox{E}\kern-.125emX}}

\begin{document}
\history{Date of publication xxxx 00, 0000, date of current version xxxx 00, 0000.}
\doi{10.1109/ACCESS.2023.0322000}

\onecolumn
\begin{framed}
   \noindent
   This work has been submitted to the IEEE for possible publication. Copyright may be transferred without notice, after which this version may no longer be accessible.%
\end{framed}
\clearpage
\twocolumn

\title{Performance of Domain-Wall Encoding in Digital Ising Machine}
\author{\uppercase{Shuta Kikuchi}\authorrefmark{1\ddag}, \uppercase{Kotaro Takahashi}\authorrefmark{1\ddag}, and \uppercase{Shu Tanaka}\authorrefmark{1, 2, 3, 4, 5}, \IEEEmembership{Member, IEEE}}

\address[1]{Graduate School of Science and Technology, Keio University, Yokohama-shi, Kanagawa 223-8522, Japan}
\address[2]{Department of Applied Physics and Physico-Informatics, Keio University, Yokohama-shi, Kanagawa 223-8522, Japan}
\address[3]{Keio University Sustainable Quantum Artificial Intelligence Center (KSQAIC), Keio University, Minato-ku, Tokyo 108-8345, Japan}
\address[4]{Human Biology-Microbiome-Quantum Research Center (WPI-Bio2Q), Keio University, Minato-ku, Tokyo 108-8345, Japan}
\address[5]{Green Computing System Research Organization, Waseda University, Shinjuku-ku, Tokyo 162-0042, Japan}
\address[\ddag]{Shuta Kikuchi and Kotaro Takahashi contributed equally to this work.}
\tfootnote{
This work was partially supported by the Japan Society for the Promotion of Science (JSPS) KAKENHI (Grant Number JP23H05447), the Council for Science, Technology, and Innovation (CSTI) through the Cross-ministerial Strategic Innovation Promotion Program (SIP), ``Promoting the application of advanced quantum technology platforms to social issues'' (Funding agency: National Institutes for Quantum Science and Technology (QST)), Japan Science and Technology Agency (JST) (Grant Number JPMJPF2221).
}

\markboth
{Kikuchi \headeretal: Performance of Domain-Wall Encoding in Digital Ising Machine}
{Kikuchi \headeretal: Performance of Domain-Wall Encoding in Digital Ising Machine}

\corresp{Corresponding author: Shuta Kikuchi (e-mail: kikuchi.shuta@keio.jp).}

\begin{abstract}
To tackle combinatorial optimization problems using an Ising machine, the objective function and constraints must be mapped onto a quadratic unconstrained binary optimization (QUBO) model.
While QUBO involves binary variables, combinatorial optimization problems frequently include integer variables, which require encoding by binary variables. 
This process, known as binary-integer encoding, includes various methods, one of which is domain-wall encoding—a recently proposed approach.
Experiments on a quantum annealing machine have demonstrated that domain-wall encoding outperforms the commonly used one-hot encoding in terms of objective function value and the probability of obtaining the optimal solution.
In a digital Ising machine, domain-wall encoding required less computation time to reach optimal solutions compared to one-hot encoding.
However, its practical effectiveness in digital Ising machines remains unclear.
To address this uncertainty, the performance of one-hot and domain-wall encoding methods was evaluated on a digital Ising machine using the quadratic knapsack problem (QKP).
The comparison focused on the dependency of penalty coefficient and sensitivity to computation time.
Domain-wall encoding demonstrated a higher feasible solution rate when relative penalty coefficients for the two constraint terms were adjusted, a strategy not commonly used in previous studies.
Additionally, domain-wall encoding obtained higher performance practical evaluation metrics for QKPs with large knapsack capacities compared to one-hot encoding.
Furthermore, it was observed to be more sensitive to computation time than one-hot encoding.
\end{abstract}

\begin{keywords}
Binary-integer encoding, combinatorial optimization problem, Ising machine, quadratic knapsack problem, quadratic unconstrained binary optimization.
\end{keywords}

\titlepgskip=-21pt

\maketitle

\section{Introduction}
\label{sec:introduction}
\PARstart{I}{sing} machines are emerging as efficient solvers for combinatorial optimization problems, which involve identifying a set of decision variables that minimize or maximize an objective function while adhering to specific constraints.
The complexity of these problems arises from the exponential growth of potential solutions with the number of decision variables. Many combinatorial optimization problems are classified as nondeterministic polynomial-time (NP)-complete or NP-hard~\cite{Karp1972reducibility}.
Such problems are pervasive across diverse social, industrial, and research domains. 
Ising machines have been successfully applied in various areas~\cite{yarkoni2022quantum}, including material simulation~\cite{king2018observation, harris2018phase, utimula2021quantum, endo2022phase, sampei2023quantum}, portfolio optimization~\cite{rosenberg2016solving, tanahashi2019application, tatsumura2023real}, traffic optimization~\cite{neukart2017traffic, irie2019quantum, ohzeki2019control, Bao2021-a, Bao2021-b, Mukasa2021, haba2022travel}, quantum compilation~\cite{naito2023isaaq}, and black-box optimization~\cite{kitai2020designing, izawa2022continuous, inoue2022towards, seki2022black, tucs2023quantum}.

Various types of Ising machines have been developed~\cite{mohseni2022ising}. 
Digital Ising machines~\cite{yamaoka201620k, okuyama2017ising, tsukamoto2017accelerator, aramon2019physics, goto2019combinatorial, kuroki2024classical, FixAE} utilize digital circuits, such as graphics processing unit (GPU), field programmable gate array (FPGA), or application specific integrated circuit (ASIC).
These machines often employ internal algorithms based on simulated annealing (SA)~\cite{kirkpatrick1983optimization, johnson1989optimization, johnson1991optimization}, as well as simulated quantum annealing~\cite{das2005quantum} and simulated bifurcations~\cite{goto2019combinatorial}.
Quantum annealing machines~\cite{johnson2011quantum, maezawa2019toward} are implemented using superconducting circuits and leverage quantum annealing~\cite{kadowaki1998quantum, tanaka2017quantum}.
Another class, coherent Ising machines~\cite{wang2013coherent}, employs optical parametric oscillators for their implementation.

To solve a combinatorial optimization problem on an Ising machine, the objective function and constraints must be translated into an \textit{Ising model}~\cite{Ising, nishimori2011elements} or its equivalent model, the \textit{quadratic unconstrained binary optimization (QUBO) model}~\cite{kochenberger2014unconstrained}.
The Hamiltonian of the Ising model, denoted as $\mathcal{H}_{\textrm{Ising}}$, is defined as follows:
\begin{align}
  \mathcal{H}_{\mathrm{Ising}} \left( \{ \sigma \} \right) = \sum_{1 \leq i < j \leq N} J_{i, j}\sigma_{i}\sigma_{j} + \sum_{i=1}^N h_{i}\sigma_{i},
  \label{eq:H_ising}
\end{align}
where $\{ \sigma \}$ is the set of spins, specifically $\{ \sigma \}= \{ \sigma_{1}, \sigma_{2},..., \sigma_{N} \}$, with each spin $\sigma_i \in \{+1, -1\}$.
In the right-hand side (RHS) of~\eqref{eq:H_ising}, the coefficients $J_{i,j}$ denote the interaction between spins $\sigma_{i}$ and $\sigma_{j}$, while $h_{i}$ is the local magnetic field acting on $\sigma_{i}$.
The Ising model can be transformed into the QUBO model through a straightforward change of variables, $x_{i} = \left( \sigma_{i} + 1 \right) / 2$, where $x_{i} \in \{0, 1\}$~\cite{tanahashi2019application, zaman2021pyqubo}.
The Hamiltonian of the QUBO model, $\mathcal{H}_{\textrm{QUBO}}$, is similarly defined by:
\begin{align}
  \mathcal{H}_{\mathrm{QUBO}}  \left( \{ x \} \right) = \sum_{1\leq i \leq j \leq N}Q_{i, j}x_{i}x_{j},
  \label{eq:H_QUBO}
\end{align}
where $\{ x \} = \{ x_{1}, x_{2},..., x_{N} \}$ and $Q_{i, j}$ is the ($i$, $j$)-th element of the $N$-by-$N$ QUBO matrix $Q$.
The off-diagonal elements of $Q$ correspond to the quadratic coefficients, while the diagonal elements represent the linear coefficients. 
We note that $x_{i}^2 = x_{i}$ for any $i$.
An Ising machine searches for the ground state (lowest-energy state) of the Ising model or QUBO. 
The configuration of variables in the ground state provides an optimal solution to the combinatorial optimization problem.

Both the Ising model and QUBO consist of binary variables. 
When combinatorial optimization problems involve integer variables, these must be encoded using binary variables.
Various methods for this encoding, collectively referred to as binary-integer encoding, have been proposed~\cite{rosenberg2016solving, chancellor2019domain, jimbo2022hybrid, dominguez2023encoding, bontekoe2023translating}.
Experimental results have shown that the performance of Ising machines depends significantly on the chosen encoding method~\cite{chancellor2019domain, chen2021performance, tamura2021performance, jimbo2022hybrid, ohno2024toward, codognet2024comparing}.
Domain-wall encoding, a recently proposed binary-integer encoding~\cite{chancellor2019domain}, constructs a problem with fewer interactions between variables compared to the widely used one-hot encoding~\cite{chancellor2019domain, chen2021performance, codognet2022domain}.
Experiments on a quantum annealing machine have demonstrated that domain-wall encoding achieves better performance than one-hot encoding in terms of objective function value and the probability of obtaining the optimal solution~\cite{chen2021performance}.
Additionally, on a digital Ising machine, domain-wall encoding showed shorter computation times required to reach optimal solutions compared to one-hot encoding~\cite{codognet2024comparing}.
However, to date, no studies have evaluated the performance of domain-wall encoding on digital Ising machines from the critical perspective of solution accuracy.
Digital Ising machines capable of handling large-scale problems with dense QUBO matrices offer an opportunity to investigate the inherent performance of domain-wall encoding.
In addition, because digital Ising machines differ in principles and specifications from quantum annealing machines, it is important to evaluate their performance independently.

This study conducts a comparative analysis of one-hot and domain-wall encoding on a digital Ising machine, specifically the Fixstars Amplify Annealing Engine (AE)~\cite{FixAE}.
The primary objective is to elucidate the performance of domain-wall encoding on digital Ising machines.
To achieve this, we compared the performance of domain-wall encoding to one-hot encoding, which requires a similar number of binary variables.
The QKP, involving integer variables, was used for this analysis.
This paper examined the dependency of penalty coefficients and sensitivity to computation time for both encoding methods.
Results indicated that there exists a specific penalty coefficient value that maximizes the performance of domain-wall encoding. 
In addition, domain-wall encoding demonstrated higher practical evaluation metrics than one-hot encoding for QKPs with large knapsack capacities. 
Furthermore, domain-wall encoding was found to be more sensitive to computation time, likely due to the absence of energy barriers for transitions between integers in this encoding scheme.

The rest of the paper is organized as follows:
Section~\ref{sec:binary_integer_cncoding} reviews binary-integer encoding methods.
Section~\ref{sec:QUBO_QKP} outlines the formulation of the QKP as a QUBO model using binary-integer encodings.
Section~\ref{sec:setup} describes the setup of the numerical experiments.
Section~\ref{sec:results} presents the results of the experiments. Section~\ref{sec:discussion} discusses the findings.
Section~\ref{sec:conclusion} concludes the paper.

\section{Binary-integer encoding methods}
\label{sec:binary_integer_cncoding}
\begin{figure}[t]
  \centering
  \subfigure[]{
    \includegraphics[clip,width=0.45\linewidth]{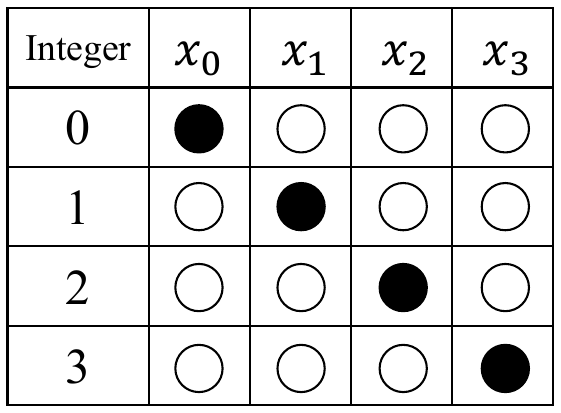}
  }
  \subfigure[]{
    \includegraphics[clip,width=0.45\linewidth]{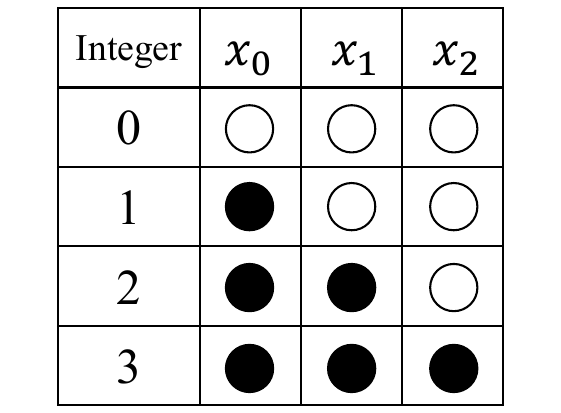}
  }
  \caption{
  Schematic of binary-integer encodings when value of integer $I \in [0, 3]$. White and black circles denote $0$ and $1$, respectively. (a) One-hot encoding and (b) domain-wall encoding.
  }
  \label{fig:schematic_encodings}
\end{figure}

In this study, we compared the performance of two binary-integer encoding methods: one-hot encoding and domain-wall encoding.
An illustrative example of these encodings is shown in Fig.~\ref{fig:schematic_encodings}.

One-hot encoding is a traditional method for binary-integer encoding.
This approach requires $(K + 1)$ binary variables to represent $(K + 1)$ integer variables.
In one-hot encoding, an integer is represented by the position of a single binary variable that takes a value of $1$.
An integer variables $I \in [0, K]$ is given by:
\begin{align}
  I = \sum_{k = 0}^{K} k x_{k}.
  \label{eq:OH_integer}
\end{align}
The one-hot encoding is enforced through energy penalization using a penalty term, denoted as $\mathcal{C}_{\mathrm{oh}}$, which is incorporated into the Hamiltonian:
\begin{align}
  \mathcal{C}_{\mathrm{oh}} = \left( \sum_{k = 0}^{K} x_{k} - 1 \right) ^ {2}.
  \label{eq:OH_encoding}
\end{align}
The $\mathcal{C}_{\mathrm{oh}}$ is minimized when only one $x_{k}$ is nonzero, ensuring the validity of the one-hot encoding.

In contrast, domain-wall encoding requires $K$ binary variables to express $(K + 1)$ integer variables.
An integer is expressed by the count of binary variables set to $1$, followed by zeros.
The boundary position between the binary variables set to $1$ and those set to $0$ is referred to as a domain wall.
An integer variable $I \in [0, K]$ is expressed by:
\begin{align}
  I = \sum_{k = 0}^{K - 1} x_{k}.
  \label{eq:DW_integer}
\end{align}
The domain wall is enforced through a penalty term, $\mathcal{C}_{\mathrm{dw}}$, incorporated into the penalty Hamiltonian, which is expressed as follows:
\begin{align}
  \mathcal{C}_{\mathrm{dw}} =  \sum_{k = 0}^{K - 2} x_{k + 1} \left( 1 - x_{k} \right).
  \label{eq:OH_encoding}
\end{align}
The $\mathcal{C}_{\mathrm{dw}}$ is minimized when exactly one domain wall exists, ensuring the validity of the domain-wall encoding.
\section{QUBO formulation for QKP}
\label{sec:QUBO_QKP}
This section presents a review of QKP and its formulation as a QUBO model utilizing binary-integer encodings.

The QKP is an extension of the well-known $0$-$1$ knapsack problem (KP), a combinatorial optimization problem defined by an inequality constraint.
The KP aims to maximize the total profit $P$ of items selected for inclusion in a knapsack, subject to the constraint that the total weight does not exceed the knapsack capacity $C$.
Each item $i$ is characterized by a weight $w_i > 0$ and a profit $p_i \geq 0$.
In the QKP, additional pairwise interaction profits $p_{i,j} \geq 0$ are introduced for each pair of items $i, j$ $(i < j)$.
For $N$ items, the QKP can be formulated as follows:
\begin{align}
    \text{Maximize} \quad & \sum_{1\leq i \leq j \leq N} p_{i,j} x_i x_j 
    \label{eq:maximize} \\
    \text{subject to} \quad & \sum_{i=1}^{N} w_i x_i \leq C
    \label{eq:subject}
\end{align}
where $x_i \in \{0, 1\}$ is a binary variable indicating whether item $i$ is selected for inclusion in the knapsack.
If item $i$ is selected, variable $x_i$ is set to a value of $1$.
Otherwise, it is set to $0$.
When both items $i$ and $j$ are selected, a pairwise profit $p_{i,j}$ is achieved.
If item $i$ is selected alone, a profit $p_{i,i}$ (i.e., $p_{i}$) is achieved.

The Hamiltonian of the QUBO for QKP, $\mathcal{H}_\mathrm{QKP}$, comprises the objective function $\mathcal{H}_\mathrm{objective}$ and the inequality constraint $\mathcal{H}_\mathrm{constraint}$. 
It is expressed as follows:
\begin{align}
  \mathcal{H}_{\mathrm{QKP}} = \mathcal{H}_\mathrm{objective} + \mu \mathcal{H}_\mathrm{constraint},
  \label{eq:H_QKP}
\end{align}
where $\mu$ is a positive constant and serves as a hyperparameter to regulate the strength of the constraint effect.

Since the Ising machine is designed to solve minimization problems, the objective function is modified by incorporating a negative sign to the original objective function.
The Hamiltonian for the objective function is expressed as follows:
\begin{align}
  \mathcal{H}_\mathrm{objective} = - \sum_{1\leq i \leq j \leq N} p_{i,j} x_i x_j.
  \label{eq:H_objective}
\end{align}

To express the constraints outlined in~\eqref{eq:subject} within the QUBO framework, it is essential to incorporate both the weight constraint and the encoding constraint.
These are denoted by $\mathcal{H}_\mathrm{weight}$ and $\mathcal{H}_\mathrm{encoding}$, respectively.
The Hamiltonian corresponding to the weight constraint, $\mathcal{H}_\mathrm{weight}$, is defined as follows:
\begin{align}
  \mathcal{H}_\mathrm{weight} = \left( \sum_{i=1}^{N} w_i x_i - C + I \right) ^ {2},
  \label{eq:H_weight}
\end{align}
where $I$ is an integer variables in the range of $[0, C]$.
The Hamiltonian for the encoding constraints, $\mathcal{H}_\mathrm{encoding}$, serves as the penalty term to enforce the correct representation of the integer.

According to Section~\ref{sec:binary_integer_cncoding}, the Hamiltonians for the weight and encoding constraints in the case of one-hot encoding are given as follows:
\begin{align}
  \mathcal{H}_\mathrm{weight}^\mathrm{one-hot} = \left( \sum_{i=1}^{N} w_i x_i - C + \sum_{k = 0}^{C} k y_{k} \right) ^ {2},
  \label{eq:H_weight_oh}
\end{align}
\begin{align}
  \mathcal{H}_\mathrm{encoding}^\mathrm{one-hot} = \left( \sum_{k = 0}^{C} y_{k} - 1 \right) ^ {2},
  \label{eq:H_encoding_oh}
\end{align}
where auxiliary variables $y_{k} \in \{ 0, 1 \}$ are introduced. 
Similarly, the Hamiltonians for the weight and encoding constraints in the case of domain-wall encoding are expressed as follows:
\begin{align}
  \mathcal{H}_\mathrm{weight}^\mathrm{domain-wall} = \left( \sum_{i=1}^{N} w_i x_i - C + \sum_{k = 0}^{C - 1} y_{k} \right) ^ {2}.
  \label{eq:H_weight_dw}
\end{align}
\begin{align}
  \mathcal{H}_\mathrm{encoding}^\mathrm{domain-wall} = \sum_{k = 0}^{C - 2} y_{k + 1} \left( 1 - y_{k} \right),
  \label{eq:H_encoding_dw}
\end{align}

Therefore, the Hamiltonian of the QUBO for QKP $\mathcal{H}_\mathrm{QKP}$ is constructed as follows:
%
\begin{align}
  \mathcal{H}_{\mathrm{QKP}} = \mathcal{H}_\mathrm{objective} + \mu \left[ \lambda \mathcal{H}_\mathrm{encoding} + \left( 1 - \lambda \right) \mathcal{H}_\mathrm{weight} \right],
  \label{eq:H_QKP_final}
\end{align}
where $\lambda$ is a positive constant and serves as a hyperparameter. 
In this study, the total Hamiltonian incorporates two hyperparameters, $\mu$ and $\lambda$, which are referred to as penalty coefficients.

\section{Setup of numerical experiments}
\label{sec:setup}

The dataset of QKP instances used in this analysis was generated in a previous study~\cite{billionnet2004exact}, with instance data files available in~\cite{QKP_instances}.
These instances are characterized by two parameters: the number of items $n \in \{ 100, 200, 300 \}$ and the density of the profit matrix $d \in \{ 25\%, 50\%, 75\%, 100\% \}$.
Each combination of $ \left( n, d \right)$ includes $10$ instances, except for $\left( 300, 75\% \right)$ and $\left( 300, 100\% \right)$, where instances are missing. 
The optimal solutions for all instances were known.
To ensure result comparisons across instances, the objective function $\mathcal{H}_\mathrm{objective}$ was normalized by dividing it by the maximum value of the profit matrix element $p_{i,j}$ in each instance.

We used the Fixstars Amplify Annealing Engine (AE)~\cite{FixAE}.
The Ising machine was implemented on a GPU with a capacity of $131072$ binary variables on a complete graph.
The machine’s internal algorithm is based on SA, and it provides users with the energy values during computation.
The computation time can be set by users.

\section{Results}
\label{sec:results}

To evaluate the performance of one-hot encoding and domain-wall encoding, the dependency of penalty coefficients and sensitivity to computation time were investigated.
This section presents the results of the numerical experiments conducted for this evaluation.

\subsection{Dependency of penalty coefficients }
\label{subsec:penalty_coefficients}

This subsection describes the one-hot and domain-wall encodings of the parameters $\lambda$ and $\mu$.
Performance was evaluated based on the probability of obtaining feasible solutions and the approximation ratio.
A feasible solution (FS) is defined as one that satisfies both the encoding and weight constraints.
The probability of obtaining a feasible solution is referred to as the FS rate, which is defined as follows:
\begin{align}
  \mathrm{FS\ rate} = \frac{\mathrm{Number\ of\ obtaining\ FSs}}{\mathrm{Total\ number\ of\ obtaining\ solutions}}.
  \label{eq:FS_rate}
\end{align}
The approximation ratio of the FSs is defined as follows:
\begin{multline}
  \mathrm{Approximation\ ratio} \\= \frac{\mathcal{H}_\mathrm{objective} \mathrm{\ of\  FS}}{\mathcal{H}_\mathrm{objective} \mathrm{\ of\  optimal\ solution}}.
  \label{eq:approximation_ratio}
\end{multline}

\begin{figure}[t]
  \centering
  \begin{minipage}{0.46\linewidth}
    \includegraphics[width=\linewidth]{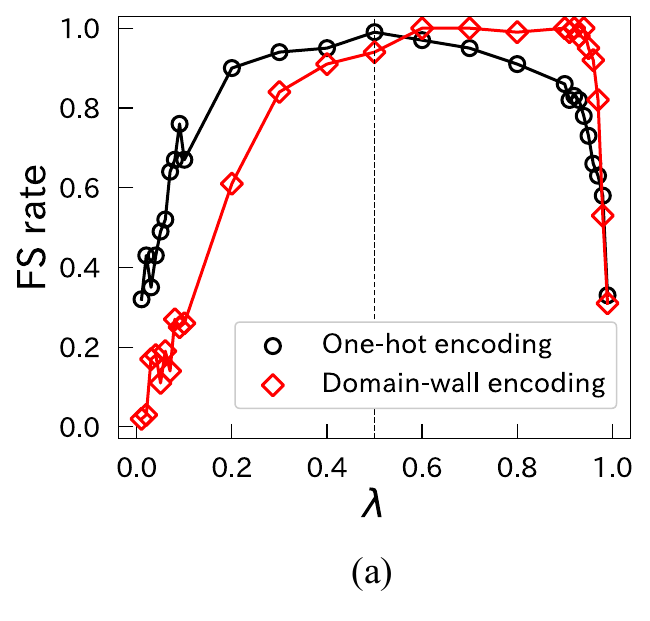}
  \end{minipage}
  \hspace{5mm}
  \begin{minipage}{0.46\linewidth}
    \includegraphics[width=\linewidth]{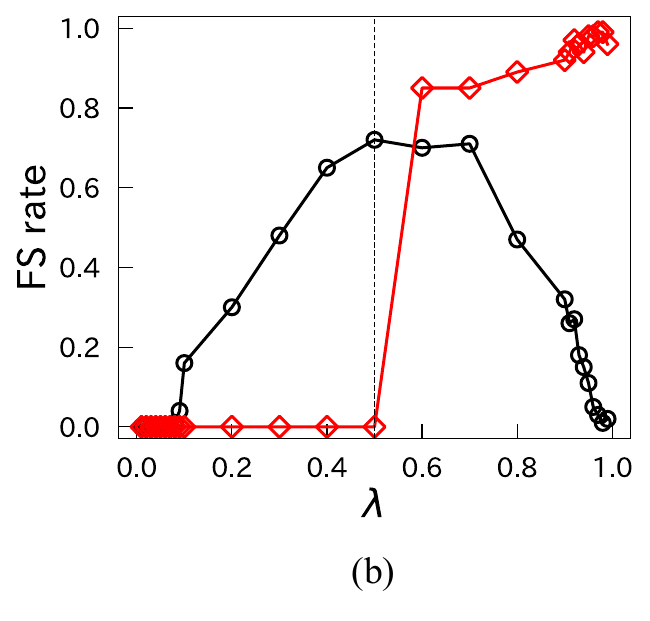}
  \end{minipage}
  \\
  \vspace{5mm}
  \begin{minipage}{0.46\linewidth}
    \includegraphics[width=\linewidth]{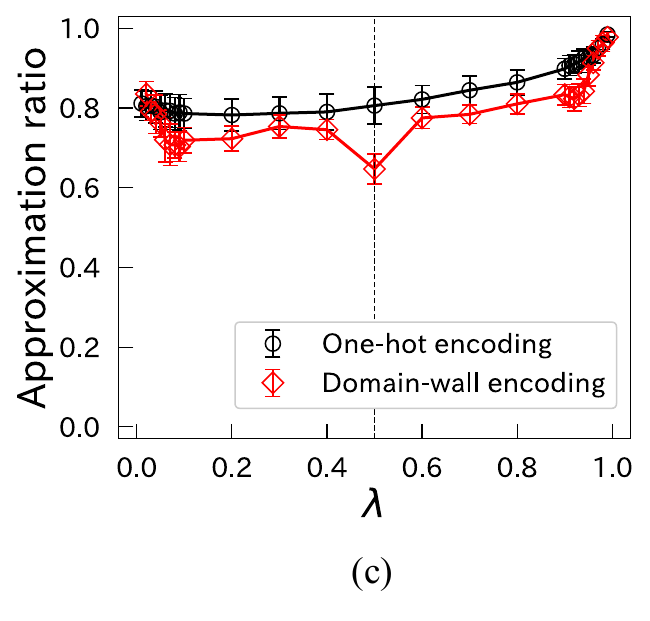}
  \end{minipage}
  \hspace{5mm}
  \begin{minipage}{0.46\linewidth}
    \includegraphics[width=\linewidth]{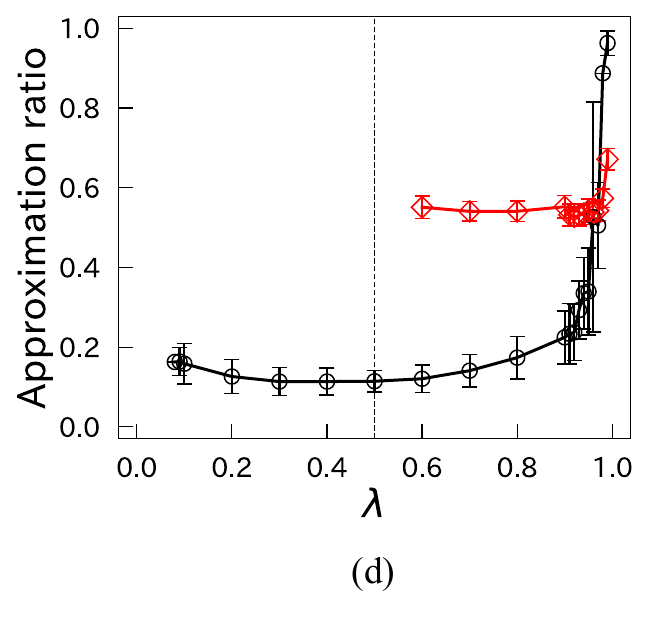}
  \end{minipage}
  \caption{$\lambda$-dependencies of one-hot encoding and domain-wall encoding under $ \left( n, d \right) = \left( 100, 25 \% \right)$. FS rates and approximate ratios are shown for instances with knapsack capacities of $669$ [(a), (c)] and $2366$ [(b), (d)]. Black circles and red diamonds denote one-hot and domain-wall encoding, respectively. Dotted vertical lines denote $\lambda = 0.5$. Each plot of approximate ratio is an average of $100$ runs and error bars are standard deviations. Solid lines between points are merely a guide to the eye.}
  \label{fig:lambda_dependencies}
\end{figure}

Figure~\ref{fig:lambda_dependencies} shows the $\lambda$-dependencies for one-hot and domain-wall encodings in QKPs with varying knapsack capacities under the conditions $ \left( n, d \right) = \left( 100, 25 \% \right)$.
Each $\lambda$ value was tested over 100 runs, with the computation time set to $10$ s and $\mu$ fixed at $10$.
The one-hot encoding exhibited a relatively high FS rate centered around $\lambda = 0.5$.
In contrast, domain-wall encoding achieved a higher FS rate for $\lambda > 0.5$.
Notably, for the instance with a knapsack capacity of $2366$, no FS was obtained at the commonly used $\lambda = 0.5$~\cite{lucas2014ising}.
For a knapsack capacity of $669$, one-hot encoding exhibited a higher approximation ratio across a wide range of $\lambda$.
In contrast, for a capacity of $2366$, domain-wall encoding demonstrated a higher approximation ratio over a broader $\lambda$ range.
Even in one-hot encoding, certain $\lambda$ values achieved a high approximation ratio, but the FS rate was low.

\begin{figure}[t]
  \centering
  \begin{minipage}{0.46\linewidth}
    \includegraphics[width=\linewidth]{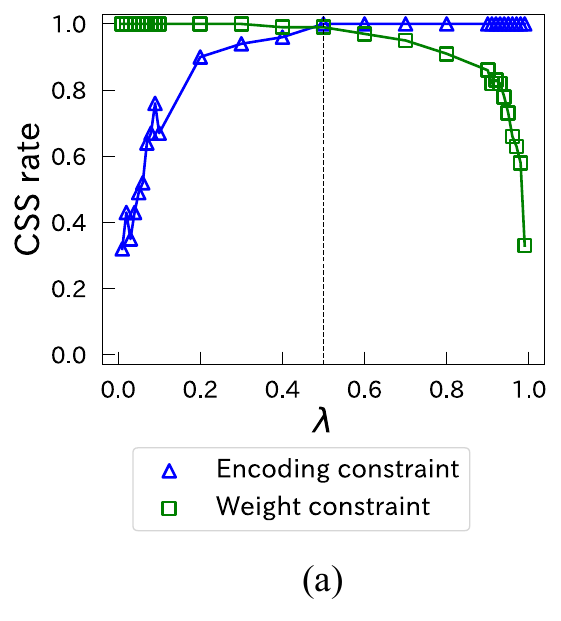}
  \end{minipage}
  \hspace{5mm}
  \begin{minipage}{0.46\linewidth}
    \includegraphics[width=\linewidth]{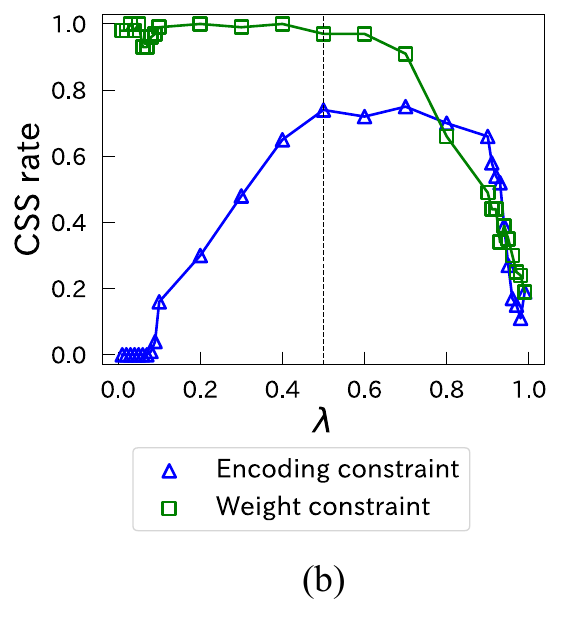}
  \end{minipage}
  \\
  \vspace{5mm}
  \begin{minipage}{0.46\linewidth}
    \includegraphics[width=\linewidth]{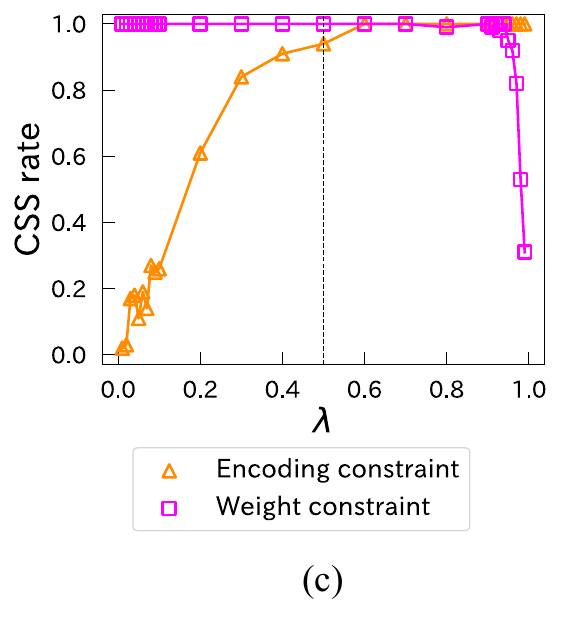}
  \end{minipage}
  \hspace{5mm}
  \begin{minipage}{0.46\linewidth}
    \includegraphics[width=\linewidth]{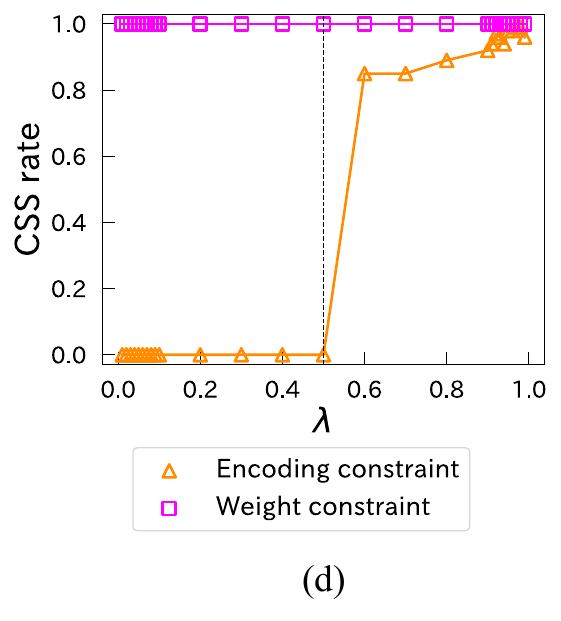}
  \end{minipage}
  \caption{$\lambda$-dependencies of CSS rates in one-hot encoding [(a), (b)] and domain-wall encoding [(c), (d)]. CSS rates are shown for instances with knapsack capacities of $669$ [(a), (c)] and $2366$ [(b), (d)] under $ \left( n, d \right) = \left( 100, 25 \% \right)$. Blue triangles and green squares denote encoding and weight constraints of one-hot encoding, respectively. Orange triangles and magenta squares denote encoding and weight constraints of domain-wall encoding, respectively. Dotted vertical lines denote $\lambda = 0.5$. Each plot of approximate ratio is an average of $100$ runs, and error bars are standard deviations. Solid lines between points are merely a guide to the eye.}
  \label{fig:lambda_dependencies_CCS}
\end{figure}

Figure~\ref{fig:lambda_dependencies_CCS} presents the constraint-satisfying solution (CSS) rates for the encoding and weight constraints.
CSSs are divided into two types: those satisfying the encoding constraint $\mathrm{CSS}_\mathrm{encoding}$ and those satisfying the weight constraint $\mathrm{CSS}_\mathrm{weight}$.
The CSS rates are defined as follows:
\begin{multline}
  \mathrm{CSS\ rate\ for\ the\ encoding\ constraint} \\= \frac{\mathrm{Number\ of\ obtaining\ \mathrm{CSSs}_\mathrm{encoding}}}{\mathrm{Total\ number\ of\ obtaining\ solutions}},
  \label{eq:CSS_rate_encoding}
\end{multline}
\begin{multline}
  \mathrm{CSS\ rate\ for\ the\ weight\ constraint} \\= \frac{\mathrm{Number\ of\ obtaining\ \mathrm{CSSs}_\mathrm{weight}}}{\mathrm{Total\ number\ of\ obtaining\ solutions}}.
  \label{eq:CSS_rate_weight}
\end{multline}

In the instance with a knapsack capacity of $669$, the CSS rate for the encoding constraint increased as $\lambda$ increased, whereas the CSS rate for the weight constraint decreased.
However, the $\lambda$ values at which the CSS rate for the weight constraint began to decrease varied between the two encodings.
For a knapsack capacity of $2366$, the CSS rate for the encoding constraint in one-hot encoding started to decline when $\lambda > 0.9$.
In contrast, domain-wall encoding maintained CSSs for the weight constraint across all values of $\lambda$. 
In addition, for domain-wall encoding, CSSs for the encoding constraint were obtained when $\lambda > 0.5$.

Practically, conditions with both a high FS rate and a high approximation ratio are considered preferable.
Consequently, the following new evaluation metric was introduced:
\begin{align}
  \mathrm{FS\ rate} \times \mathrm{approximation\ ratio}.
  \label{eq:product_of_FSrate_approximationratio}
\end{align}

\begin{figure}[t]
  \centering
  \begin{minipage}{0.46\linewidth}
    \includegraphics[width=\linewidth]{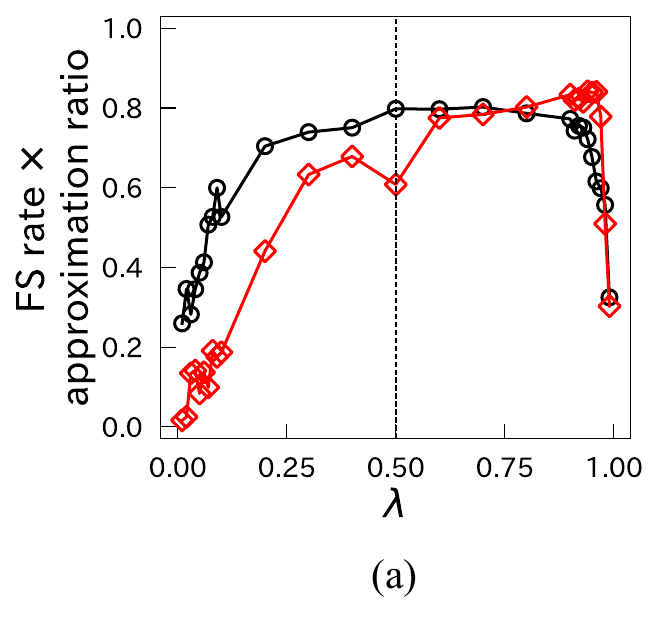}
  \end{minipage}
  \hspace{5mm}
  \begin{minipage}{0.46\linewidth}
    \includegraphics[width=\linewidth]{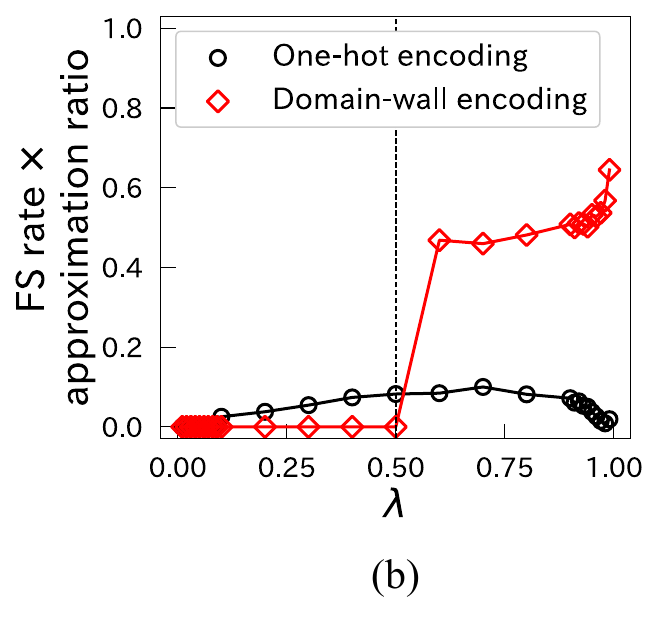}
  \end{minipage}
  \caption{$\lambda$-dependencies of FS rate $\times$ approximation ratio~\eqref{eq:product_of_FSrate_approximationratio} for instances with knapsack capacities of (a) $669$ and (b) $2366$ under $ \left( n, d \right) = \left( 100, 25 \% \right)$. Black circles and red diamonds denote one-hot encoding and domain-wall encoding, respectively. Dotted vertical lines denote $\lambda = 0.5$. Solid lines between points are merely a guide to the eye.}
  \label{fig:lambda_dependencies_new}
\end{figure}

Figure~\ref{fig:lambda_dependencies_new} shows the $\lambda$-dependencies of the new evaluation metric defined in~\eqref{eq:product_of_FSrate_approximationratio}.
The domain-wall encoding exhibited higher values compared to one-hot encoding in regions where $\lambda$ was large.

The new evaluation metric~\eqref{eq:product_of_FSrate_approximationratio} was applied to instances with varying knapsack capacities under $ \left( n, d \right) = \left( 100, 25 \% \right)$.
The value of $\lambda$ was incremented by $0.01$, ranging from $0.01$ to $0.99$, with $10$ runs performed for each setting.
Table~\ref{table:100_25} illustrates the maximum values of the new evaluation metric~\eqref{eq:product_of_FSrate_approximationratio} and the corresponding $\lambda$ values (denoted as $\lambda^{*}$) for each instance.
The results in Table~\ref{table:100_25} indicate that domain-wall encoding significantly outperforms one-hot encoding in instances with large knapsack capacities.
In addition, the maximum value of the new evaluation metric~\eqref{eq:product_of_FSrate_approximationratio} for domain-wall encoding is consistently achieved at $\lambda > 0.9$.
These trends were observed across instances with different combinations of $ \left( n, d \right)$.
Detailed results are presented in Appendix~\ref{sec:appendixA}.

\begin{table}[t]
    \centering
    \caption{Knapsack capacity and maximum value of FS rate $\times$ approximation ratio~\eqref{eq:product_of_FSrate_approximationratio} under $ \left( n, d \right) = \left( 100, 25 \% \right)$. $\lambda^{*}$ corresponds to $\lambda$ at which the maximum value of FS rate $\times$ approximation ratio is obtained.}
    \label{table:100_25}
    \begin{tabular}{cccccc}
    \toprule
    \raisebox{-1.0em}{\shortstack[c]{Knapsack \\ capacity}} & 
    \multicolumn{2}{c}{One-hot encoding} & 
    \multicolumn{2}{c}{Domain-wall encoding} \\ 
    \cmidrule(lr){2-3} \cmidrule(lr){4-5}
    & \shortstack[c]{FS rate $\times$\\ approximation ratio} & \raisebox{0.5em}{$\lambda^{*}$} 
    & \shortstack[c]{FS rate $\times$\\ approximation ratio} & \raisebox{0.5em}{$\lambda^{*}$} \\ 
    \midrule
    156 & 1.0 & 0.95 & 1.0 & 0.97 \\
    466 & 0.945 & 0.91 & 0.942 & 0.94 \\
    619 & 0.886 & 0.89 & 0.899 & 0.91 \\
    669 & 0.902 & 0.89 & 0.910 & 0.96 \\
    1040 & 0.792 & 0.83 & 0.825 & 0.98 \\
    1236 & 0.640 & 0.90 & 0.784 & 0.98 \\
    1597 & 0.446 & 0.97 & 0.834 & 0.99 \\
    2298 & 0.357 & 0.99 & 0.664 & 0.99 \\
    2366 & 0.145 & 0.89 & 0.681 & 0.99 \\
    2536 & 0.130 & 0.63 & 0.675 & 0.99 \\
    \bottomrule
    \end{tabular}
\end{table}

\begin{figure}[t]
  \centering
  \begin{minipage}{0.46\linewidth}
    \includegraphics[width=\linewidth]{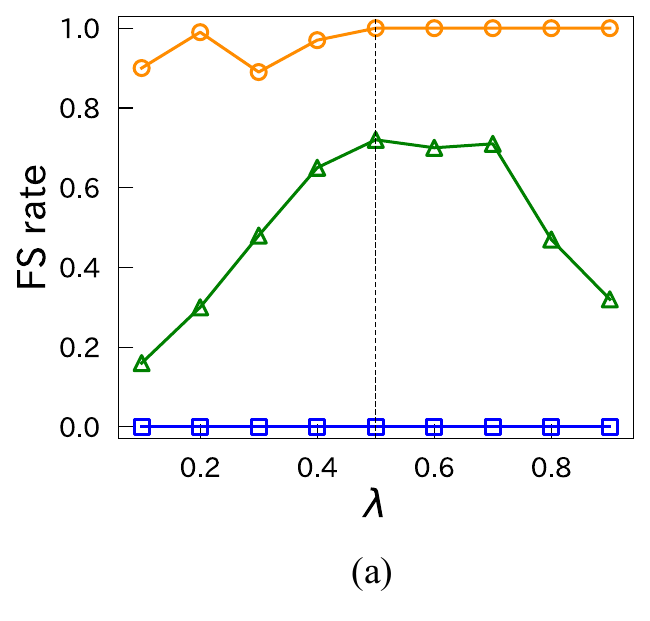}
  \end{minipage}
  \hspace{5mm}
  \begin{minipage}{0.46\linewidth}
    \includegraphics[width=\linewidth]{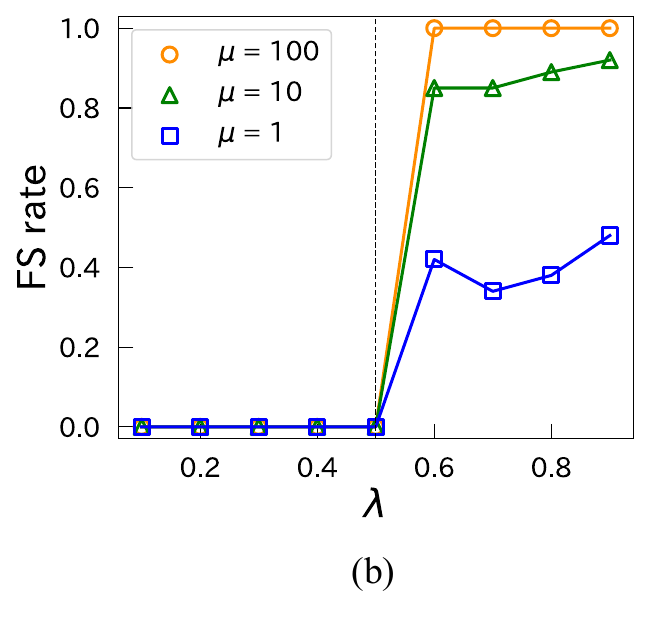}
  \end{minipage}
  \\
  \vspace{5mm}
  \begin{minipage}{0.46\linewidth}
    \includegraphics[width=\linewidth]{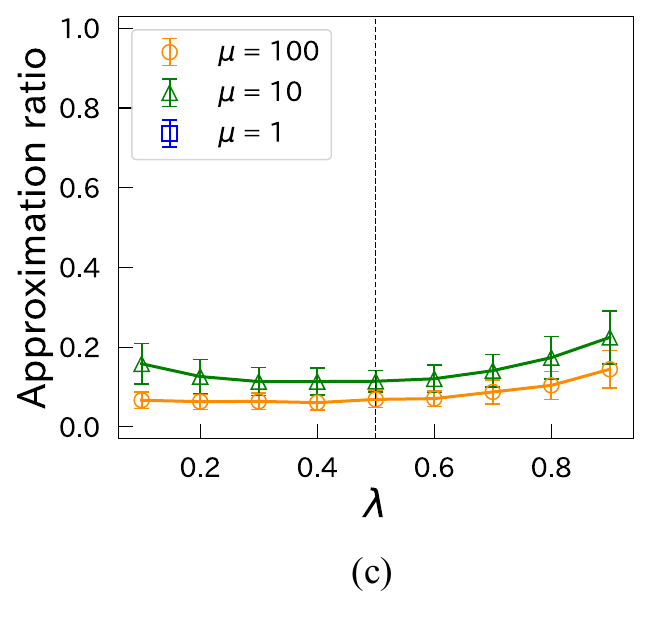}
  \end{minipage}
  \hspace{5mm}
  \begin{minipage}{0.46\linewidth}
    \includegraphics[width=\linewidth]{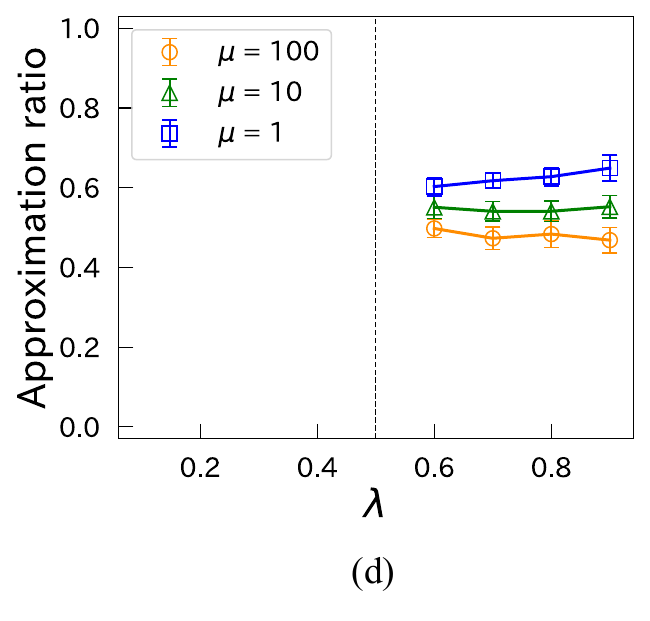}
  \end{minipage}
  \caption{$\mu$-dependencies of each encoding for QKP with knapsack capacity of $2366$ under $ \left( n, d \right) = \left( 100, 25 \% \right)$. FS rates and approximation ratios are shown for one-hot encoding [(a), (c)] and domain-wall encoding [(b), (d)]. Blue squares, green triangles, and orange circles denote $\mu = 1, 10, 100$, respectively, for one-hot and domain-wall encoding. Dotted lines denote $\lambda = 0.5$. Every plot of approximation ratio is an average of $100$ runs. Error bars are standard deviations. Solid lines between points are merely a guide to the eye.}
  \label{fig:mu_dependencies}
\end{figure}

An evaluation was conducted to determine the dependency of each encoding on the variable $\mu$.
The value of $\lambda$ was varied in increments of $0.1$ ranging from $0.1$ to $0.9$.
One hundred runs were performed for each $\lambda$ setting, with a computation time of $10$ s per run.
Figure~\ref{fig:mu_dependencies} shows the $\mu$-dependencies for a QKP with a knapsack capacity of $2366$ under $ \left( n, d \right) = \left( 100, 25 \% \right)$. 
As $\mu$ increased, the FS rate rose while the approximation ratio declined for both encoding methods.
Furthermore, domain-wall encoding obtained feasible solutions even at smaller values of $\mu$ ($\mu = 1$) compared to one-hot encoding.

\subsection{Sensitivity to computation time}
\label{subsec:timeout}

\begin{figure}[t]
  \centering
  \begin{minipage}{0.46\linewidth}
    \includegraphics[width=\linewidth]{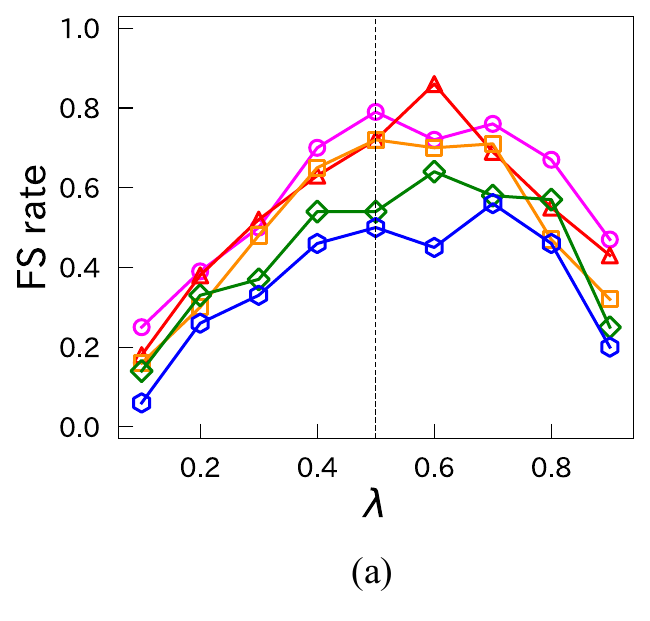}
  \end{minipage}
  \hspace{5mm}
  \begin{minipage}{0.46\linewidth}
    \includegraphics[width=\linewidth]{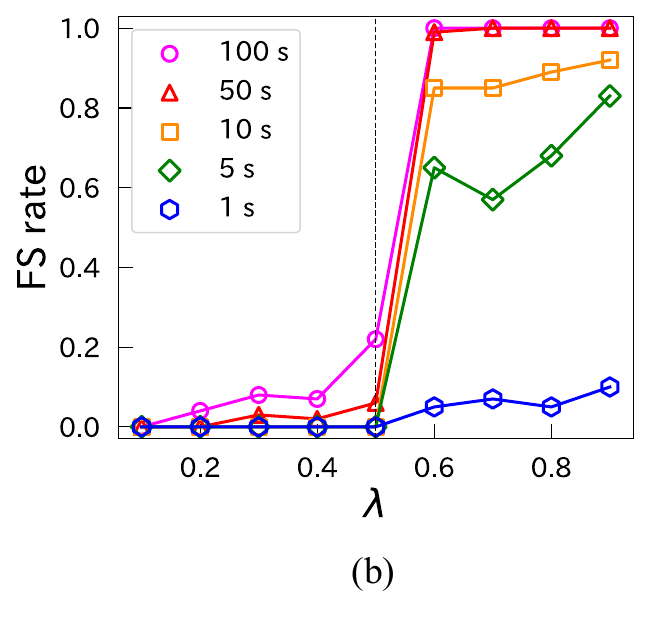}
  \end{minipage}
  \\
  \vspace{5mm}
  \begin{minipage}{0.46\linewidth}
    \includegraphics[width=\linewidth]{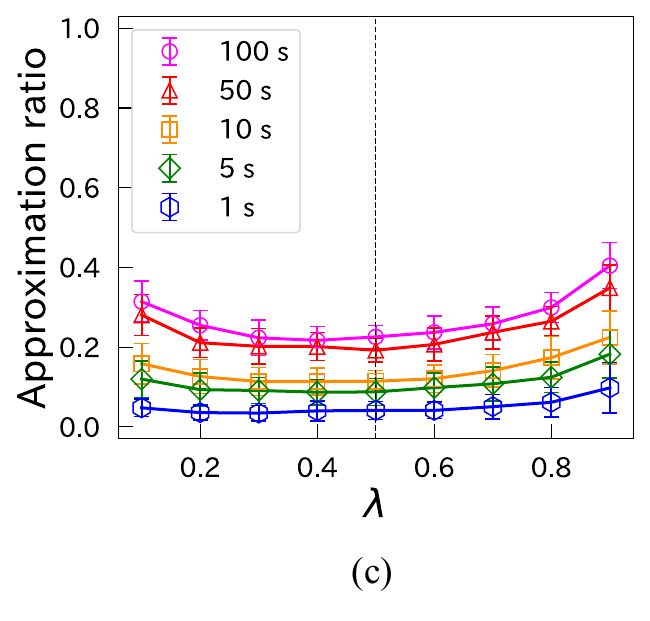}
  \end{minipage}
  \hspace{5mm}
  \begin{minipage}{0.46\linewidth}
    \includegraphics[width=\linewidth]{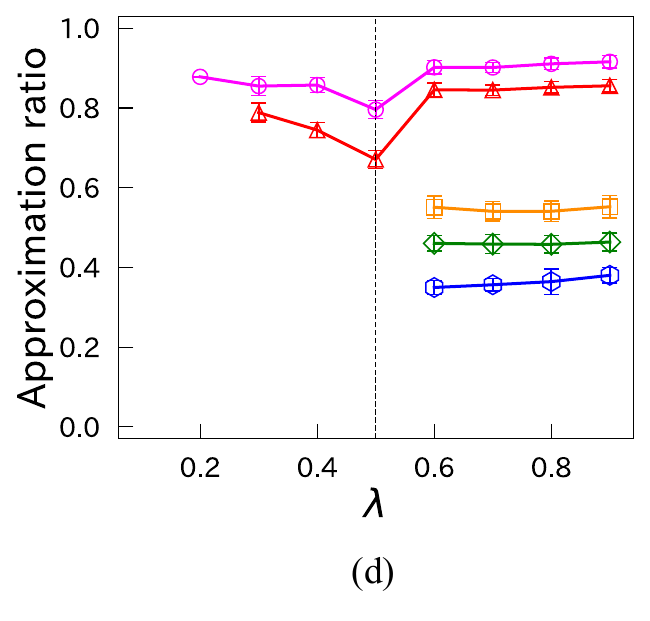}
  \end{minipage}
  \caption{Sensitivity to computation time of each encoding. FS rates and approximation ratios are shown for one-hot encoding [(a), (c)] and domain-wall encoding [(b), (d)]. Blue hexagons, green diamonds, orange squares, red triangles, and magenta circles represent computation times of $1, 5, 10, 50$, and $100$ s, respectively, for one-hot and domain-wall encoding. Dotted vertical lines denote $\lambda = 0.5$. Every plot of approximation ratio is an average of $100$ runs. Error bars are standard deviations. Solid lines between points are merely a guide to the eye.}
  \label{fig:100_25_2_timeout}
\end{figure}

In this subsection, we assessed the sensitivity of each encoding to computation time.
Computation times were set to $1, 5, 10, 50$, and $100$ s.
The value of $\mu$ was fixed at $10$,
and $\lambda$ was varied between $0.1$ and $0.9$ in increments of $0.1$, with $100$ runs performed for each setting.
Analysis was conducted using an instance with a knapsack capacity of $2366$ under $ \left( n, d \right) = \left( 100, 25 \% \right)$ as the QKP. 
Figure~\ref{fig:100_25_2_timeout} presents the FS rates and approximation ratios for various computation times. 
Both encodings showed an increase in FS rate and approximation ratio values with longer computation times.
However, domain-wall encoding exhibited a more pronounced dependency on computation time, indicating its high sensitivity to this parameter.
This trend was consistently observed across other instances as well (see Appendix~\ref{sec:appendixB}).

\begin{figure}[t]
  \centering
  \includegraphics[clip,width=0.6\linewidth]{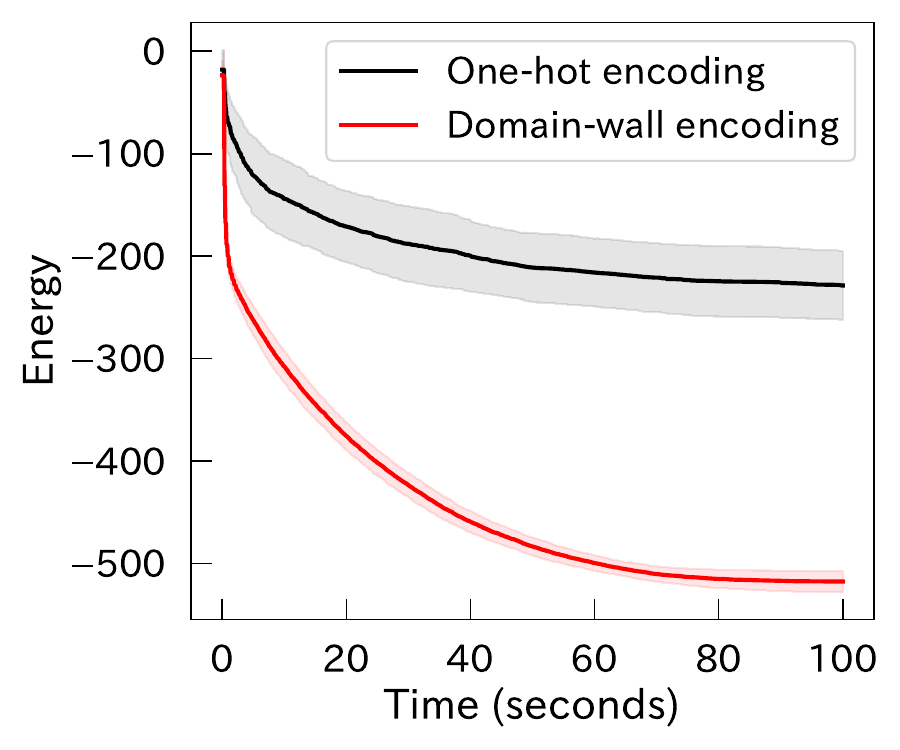}
  \caption{Dynamical process during computation. An instance with knapsack capacity of $2366$ under $ \left( n, d \right) = \left( 100, 25 \% \right)$ is used as QKP. Value of $\lambda$ and computation time were set to $0.9$ and $100$ s, respectively. Black and red lines denote one-hot and domain-wall encoding, respectively. Lines represent average of $100$ runs. Color bars are standard deviations.}
  \label{fig:dynamical_process}
\end{figure}

Next, we investigated the dynamical process occurring during computation.
Figure~\ref{fig:dynamical_process} shows the dynamical process for a computation time of $100$ s and $\lambda = 0.9$, as shown in Fig.~\ref{fig:100_25_2_timeout}.
This energy corresponds to the value of $\mathcal{H}_{\mathrm{QKP}}$.
For one-hot encoding, the energy updates decelerated early in the computation, indicating a rapid convergence. 
In contrast, for domain-wall encoding, the energy updates continued throughout the computation time.

\section{Discussion}
\label{sec:discussion}
The CSS rate for the weight constraint in domain-wall encoding remained high even for large $\lambda$, indicating that the penalty coefficient for the weight constraint was relatively small, as shown in Fig.~\ref{fig:lambda_dependencies_CCS}. 
This is attributed to the term $\sum_{k = 0}^{C - 1} y_{k}$ in $\mathcal{H}_\mathrm{weight}^\mathrm{domain-wall}$ in~\eqref{eq:H_weight_dw}, which represents the number of auxiliary variables with a value of $1$, used to express an integer.
Since this representation allows integers to be adjusted incrementally by one, it simplifies the expression of integers, thereby satisfying the constraint even with a small penalty coefficient.
As the penalty coefficient increases, the CSS rate for the encoding constraint improves steadily, leading to a higher FS rate for larger $\lambda$.
In QKP instances with large knapsack capacities, the difference between $\sum_{i=1}^{N} w_i x_i$ and $C$ in~\eqref{eq:H_weight_dw} becomes significant, allowing the CSS rate to increase even with smaller penalty coefficients. 
Consequently, the highest FS rate was observed at higher values of $\lambda$ (Fig.~\ref{fig:lambda_dependencies}).
Furthermore, a high FS rate contributes to a high value of the FS rate $\times$ approximation ratio~\eqref{eq:product_of_FSrate_approximationratio} (Fig.~\ref{fig:lambda_dependencies_new} and Table~\ref{table:100_25}).

Domain-wall encoding demonstrated the ability to obtain FSs even at lower $\mu$ values compared to one-hot encoding (Fig.~\ref{fig:mu_dependencies}). 
This behavior is also related to the term~\eqref{eq:H_weight_dw}.
A previous study showed that unary encoding could achieve feasible solutions with lower $\mu$ values compared to one-hot encoding~\cite{tamura2021performance}.
Unary encoding, which is equivalent to domain-wall encoding without $\mathcal{H}_\mathrm{encoding}^\mathrm{domain-wall}$, shares the same $\mathcal{H}_\mathrm{weight}^\mathrm{domain-wall}$ term. 
This suggests that this formulation affects the $\mu$-dependency.

\begin{figure}[t]
  \centering
  \begin{minipage}{0.8\linewidth}
    \includegraphics[width=\linewidth]{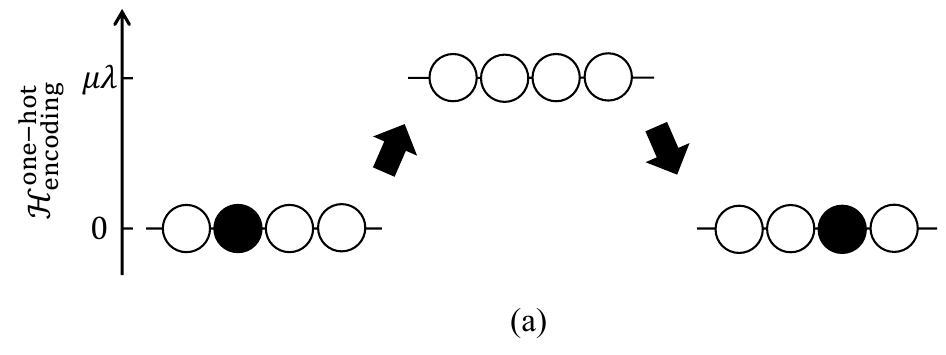}
  \end{minipage}
  \\
  \vspace{5mm}
  \begin{minipage}{0.8\linewidth}
    \includegraphics[width=\linewidth]{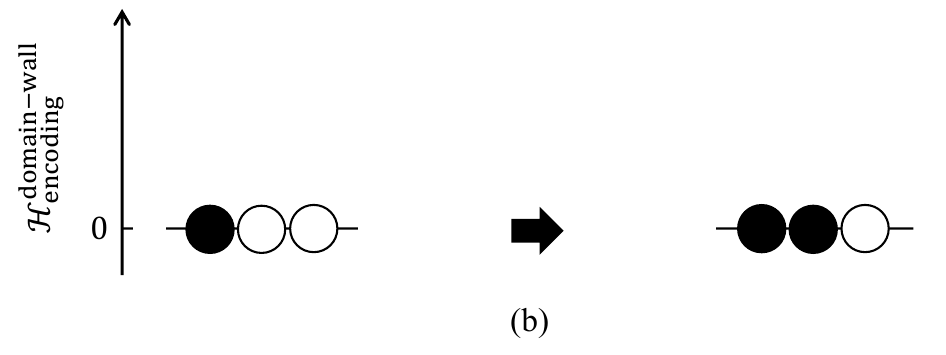}
  \end{minipage}
  \caption{
  Example of transition of $\mathcal{H}_\mathrm{encoding}$ between integers in (a) one-hot encoding and (b) domain-wall encoding. White and black circles denote $0$ and $1$, respectively.
  }
  \label{fig:energy_example}
\end{figure}

Domain-wall encoding also exhibited sensitivity to computation time.
Given that the Fixstars Amplify AE employs an SA-based algorithm, it is inferred that domain-wall encoding continues to update solutions even at low temperatures.
This behavior can be linked to the ease of transitioning between integers at low temperatures in SA.
An example is shown in Fig.~\ref{fig:energy_example}.
In one-hot encoding, transitioning between integers requires passing through an infeasible solution, where all auxiliary spins are zero~\cite{berwald2023understanding}. 
An energy barrier of $\mu \lambda$ exists between feasible and infeasible solutions, making transitions difficult at low temperatures, as solutions remain trapped in local minima. 
However, domain-wall encoding exhibits no energy barrier between neighboring integers, allowing transitions with equal energy levels.
This facilitates exploration of low-energy states, even at low temperatures.

\section{Conclusion}
\label{sec:conclusion}

The performance of one-hot encoding and domain-wall encoding in a digital Ising machine was evaluated by formulating QKPs into a QUBO model utilizing each binery-integer encoding.
The Hamiltonian of the QUBO for the QKP included two penalty coefficients, $\mu$ and $\lambda$.
To compare the performance of the two encodings, the dependency of penalty coefficients and sensitivity to computation time were investigated.
The results demonstrated that domain-wall encoding achieved a higher FS rate for $\lambda > 0.5$, even though $\lambda = 0.5$ is a commonly used value~\cite{lucas2014ising}.
In addition, domain-wall encoding exhibited superior performance in practical evaluation metrics~\eqref{eq:product_of_FSrate_approximationratio}, particularly for QKPs with large knapsack capacities.
Domain-wall encoding was also capable of obtaining feasible solutions with smaller $\mu$ compared to one-hot encoding.
Regarding computation time, domain-wall encoding was found to be more sensitive than one-hot encoding, maintaining solution updates for longer durations.

These unique characteristics of domain-wall encoding have not been reported previously.
It remains unclear whether these characteristics are specific to digital Ising machines employing SA-based algorithms.
Future work will focus on verifying whether these characteristics also appear when using quantum annealing machines and other Ising machines that employ different algorithms.

\appendices
\section{Penalty coefficient dependencies of multiple QKP instances}
\label{sec:appendixA}

This appendix provides an analysis of the $\lambda$-dependencies across multiple QKP instances.
Table~\ref{table:100_50_75_100} lists the maximum values of the new evaluation metric~\eqref{eq:product_of_FSrate_approximationratio} and the corresponding $\lambda$ ($\lambda^{*}$) for instances with $ \left( n, d \right) = \left( 100, 50 \% \right)$, $\left( 100, 75 \% \right)$, and $\left( 100, 100 \% \right)$.
In addition, Tables~\ref{table:200_25_50_75_100} and~\ref{table:300_25_50} illustrate results for instances with $\left( 200, 25 \% \right)$, $\left( 200, 50 \% \right)$, $\left( 200, 75 \% \right)$, $\left( 200, 100 \% \right)$, $\left( 300, 25 \% \right)$, and $\left( 300, 50 \% \right)$.
The $\lambda$ value was incremented by $0.01$, ranging from $0.01$ to $0.99$, with $10$ runs performed for each setting.
Computation time was set to $10$ s, and $\mu$ was fixed at $10$.

The results in Tables~\ref{table:100_25},~\ref{table:100_50_75_100},~\ref{table:200_25_50_75_100}, and~\ref{table:300_25_50} consistently show that domain-wall encoding significantly outperforms one-hot encoding in instances with large knapsack capacities.
In addition, the maximum values of the new evaluation metric~\eqref{eq:product_of_FSrate_approximationratio} for domain-wall encoding were consistently attained when $\lambda > 0.9$.

In two instances from Table~\ref{table:300_25_50}, where the knapsack capacities exceeded $6000$, no FS was obtained using domain-wall encoding, while one FS was obtained with one-hot encoding.
This discrepancy likely results from the large difference between $\sum_{i=1}^{N} w_i x_i$ and $C$ in~\eqref{eq:H_weight_dw}, which reduced the relative effectiveness of the penalty coefficient in $\mathcal{H}_\mathrm{encoding}^\mathrm{domain-wall}$.
To address this, simulations were conducted with $\lambda$ incremented by $0.001$ within the range of $0.991$ to $0.999$.
Consequently, domain-wall encoding successfully obtained multiple FSs for each instance.
For instances with knapsack capacities of $6282$ and $7040$, the maximum values of the FS rate $\times$ approximation ratio~\eqref{eq:product_of_FSrate_approximationratio} were $0.385$ ($\lambda^{*}$=$0.994$) and $0.592$ ($\lambda^{*}$=$0.992$), respectively, both surpassing the performance of one-hot encoding.

\begin{table*}[t]
    \centering
    \caption{Knapsack capacity and maximum value of FS rate $\times$ approximation ratio\eqref{eq:product_of_FSrate_approximationratio} with instances of $100$ items. $\lambda^{*}$ corresponds to $\lambda$ at which maximum value of FS rate $\times$ approximation ratio is obtained.}
    \label{table:100_50_75_100}
    \begin{tabular}{ccccccc}
    \toprule
    \raisebox{-1.0em}{\shortstack[c]{The number of \\ items}} &
    \raisebox{-1.0em}{\shortstack[c]{Density of \\ the profit matrix}} &
    \raisebox{-1.0em}{\shortstack[c]{Knapsack \\ capacity}} & 
    \multicolumn{2}{c}{One-hot encoding} & 
    \multicolumn{2}{c}{Domain-wall encoding} \\ 
    \cmidrule(lr){4-5} \cmidrule(lr){6-7}
    & & & \shortstack[c]{FS rate $\times$\\ approximation ratio} & \raisebox{0.5em}{$\lambda^{*}$} 
    & \shortstack[c]{FS rate $\times$\\ approximation ratio} & \raisebox{0.5em}{$\lambda^{*}$} \\ 
    \midrule
    100 & 50 & 466 & 0.799 & 0.94 & 0.806 & 0.97 \\
     &  & 568 & 0.924 & 0.87 & 0.870 & 0.85 \\
     &  & 619 & 0.776 & 0.91 & 0.780 & 0.96 \\
     &  & 983 & 0.943 & 0.97 & 0.753 & 0.96 \\
     &  & 1236 & 1.0 & 0.98 & 1.0 & 0.99 \\
     &  & 1459 & 0.591 & 0.91 & 0.695 & 0.97 \\
     &  & 1562 & 0.644 & 0.97 & 0.782 & 0.98 \\
     &  & 1918 & 0.314 & 0.94 & 0.862 & 0.99 \\
     &  & 2019 & 0.184 & 0.98 & 0.864 & 0.99 \\
     &  & 2208 & 0.235 & 0.99 & 0.654 & 0.99 \\

    \midrule
    100 & 75 & 370 & 1.0 & 0.98 & 0.925 & 0.88 \\
     &  & 382 & 0.980 & 0.93 & 0.912 & 0.93 \\
     &  & 847 & 0.869 & 0.88 & 0.912 & 0.96 \\
     &  & 953 & 0.892 & 0.85 & 0.883 & 0.95 \\
     &  & 1346 & 0.731 & 0.94 & 0.799 & 0.96 \\
     &  & 1498 & 0.646 & 0.95 & 0.895 & 0.98 \\
     &  & 1542 & 0.582 & 0.98 & 0.742 & 0.98 \\
     &  & 1918 & 0.457 & 0.98 & 0.754 & 0.98 \\
     &  & 1924 & 0.283 & 0.97 & 0.895 & 0.98 \\
     &  & 2466 & 0.104 & 0.60 & 0.892 & 0.99 \\

     \midrule
     100 & 100 & 100 & 1.0 & 0.13 & 1.0 & 0.85 \\
     &  & 574 & 0.920 & 0.79 & 0.874 & 0.81 \\
     &  & 608 & 1.0 & 0.98 & 0.881 & 0.88 \\
     &  & 616 & 0.897 & 0.71 & 0.896 & 0.85 \\
     &  & 726 & 0.908 & 0.84 & 0.878 & 0.93 \\
     &  & 1779 & 0.497 & 0.98 & 0.885 & 0.97 \\
     &  & 1978 & 0.295 & 0.98 & 0.809 & 0.98 \\
     &  & 2232 & 0.192 & 0.97 & 0.653 & 0.98 \\
     &  & 2257 & 0.178 & 0.98 & 0.754 & 0.98 \\
     &  & 2319 & 0.060 & 0.65 & 0.883 & 0.99 \\    
    \bottomrule
    \end{tabular}
\end{table*}

\begin{table*}[t]
    \centering
    \caption{Knapsack capacity and maximum value of FS rate $\times$ approximation ratio~\eqref{eq:product_of_FSrate_approximationratio} with instances of $200$ items. $\lambda^{*}$ corresponds to $\lambda$ at which maximum value of FS rate $\times$  approximation ratio is obtained.}
    \label{table:200_25_50_75_100}
    \begin{tabular}{ccccccc}
    \toprule
    \raisebox{-1.0em}{\shortstack[c]{The number of \\ items}} &
    \raisebox{-1.0em}{\shortstack[c]{Density of \\ the profit matrix}} &
    \raisebox{-1.0em}{\shortstack[c]{Knapsack \\ capacity}} & 
    \multicolumn{2}{c}{One-hot encoding} & 
    \multicolumn{2}{c}{Domain-wall encoding} \\ 
    \cmidrule(lr){4-5} \cmidrule(lr){6-7}
    & & & \shortstack[c]{FS rate $\times$\\  approximation ratio} & \raisebox{0.5em}{$\lambda^{*}$} 
    & \shortstack[c]{FS rate $\times$\\  approximation ratio} & \raisebox{0.5em}{$\lambda^{*}$} \\ 
    \midrule
    200 &  25  & 990  & 0.738 & 0.55 & 0.712 & 0.94 \\
    &    & 1012 & 0.773 & 0.74 & 0.825 & 0.95 \\
    &    & 1242 & 0.718 & 0.60 & 0.798 & 0.96 \\
    &    & 1700 & 0.721 & 0.82 & 0.716 & 0.96 \\
    &    & 2861 & 0.185 & 0.99 & 0.685 & 0.98 \\
    &    & 3688 & 0.056 & 0.70 & 0.383 & 0.99 \\
    &    & 4027 & 0.085 & 0.90 & 0.224 & 0.99 \\
    &    & 4271 & 0.045 & 0.73 & 0.149 & 0.99 \\
    &    & 4518 & 0.039 & 0.80 & 0.111 & 0.91 \\
    &    & 4763 & 0.043 & 0.81 & 0.092 & 0.94 \\
    
    \midrule
    200 & 50 & 1105 & 0.772 & 0.70 & 0.732 & 0.93 \\
        &    & 2120 & 0.418 & 0.93 & 0.775 & 0.95 \\
        &    & 2132 & 0.354 & 0.96 & 0.667 & 0.93 \\
        &    & 2245 & 0.400 & 0.91 & 0.688 & 0.97 \\
        &    & 2430 & 0.342 & 0.96 & 0.812 & 0.97 \\
        &    & 2756 & 0.355 & 0.96 & 0.710 & 0.98 \\
        &    & 3159 & 0.091 & 0.97 & 0.321 & 0.98 \\
        &    & 3547 & 0.0   & -    & 0.288 & 0.99 \\
        &    & 4354 & 0.051 & 0.84 & 0.102 & 0.93 \\
        &    & 4833 & 0.020 & 0.69 & 0.053 & 0.95 \\
    
    \midrule
    200 & 75 & 408 & 0.928 & 0.78 & 0.935 & 0.86 \\
     &  & 859 & 0.817 & 0.47 & 0.801 & 0.90 \\
     &  & 955 & 0.779 & 0.70 & 0.787 & 0.90 \\
     &  & 1220 & 0.715 & 0.54 & 0.798 & 0.95 \\
     &  & 1349 & 0.681 & 0.45 & 0.809 & 0.91 \\
     &  & 1736 & 0.543 & 0.82 & 0.845 & 0.94 \\
     &  & 2139 & 0.528 & 0.95 & 0.747 & 0.96 \\
     &  & 2866 & 0.099 & 0.98 & 0.695 & 0.97 \\
     &  & 3412 & 0.091 & 0.98 & 0.825 & 0.98 \\
     &  & 3412 & 0.0 & - & 0.726 & 0.98 \\

    \midrule
    200 & 100 & 173 & 1.0 & 0.80 & 0.973 & 0.77 \\
     &  & 195 & 1.0 & 0.80 & 0.963 & 0.72 \\
     &  & 498 & 0.901 & 0.67 & 0.908 & 0.86 \\
     &  & 1325 & 0.683 & 0.85 & 0.803 & 0.93 \\
     &  & 1859 & 0.602 & 0.60 & 0.835 & 0.93 \\
     &  & 3256 & 0.289 & 0.97 & 0.614 & 0.97 \\
     &  & 3583 & 0.098 & 0.97 & 0.571 & 0.98 \\
     &  & 3677 & 0.186 & 0.98 & 0.666 & 0.98 \\
     &  & 4005 & 0.0 & - & 0.296 & 0.99 \\
     &  & 4785 & 0.0 & - & 0.100 & 0.99 \\  
    \bottomrule
    \end{tabular}
\end{table*}

\begin{table*}[t]
    \centering
    \caption{Knapsack capacity and maximum value of FS rate $\times$ approximation ratio~\eqref{eq:product_of_FSrate_approximationratio} with instances of $300$ items. $\lambda^{*}$ corresponds to $\lambda$ at which maximum value of FS rate $\times$ approximation ratio is obtained.}
    \label{table:300_25_50}
    \begin{tabular}{ccccccc}
    \toprule
    \raisebox{-1.0em}{\shortstack[c]{The number of \\ items}} &
    \raisebox{-1.0em}{\shortstack[c]{Density of \\ the profit matrix}} &
    \raisebox{-1.0em}{\shortstack[c]{Knapsack \\ capacity}} & 
    \multicolumn{2}{c}{One-hot encoding} & 
    \multicolumn{2}{c}{Domain-wall encoding} \\ 
    \cmidrule(lr){4-5} \cmidrule(lr){6-7}
    & & & \shortstack[c]{FS rate $\times$\\ approximation ratio} & \raisebox{0.5em}{$\lambda^{*}$} 
    & \shortstack[c]{FS rate $\times$\\ approximation ratio} & \raisebox{0.5em}{$\lambda^{*}$} \\ 
    \midrule
    300 & 25 & 114 & 0.997 & 0.82 & 1.0 & 0.80 \\
     &  & 197 & 0.960 & 0.81 & 0.995 & 0.88 \\
     &  & 376 & 0.883 & 0.65 & 0.919 & 0.86 \\
     &  & 3128 & 0.185 & 0.75 & 0.746 & 0.97 \\
     &  & 3150 & 0.204 & 0.85 & 0.669 & 0.97 \\
     &  & 3451 & 0.076 & 0.94 & 0.425 & 0.98 \\
     &  & 3878 & 0.191 & 0.99 & 0.357 & 0.93 \\
     &  & 5322 & 0.024 & 0.75 & 0.180 & 0.99 \\
     &  & 5759 & 0.026 & 0.78 & 0.043 & 0.90 \\
     &  & 6282 & 0.012 & 0.85 & 0.0 & - \\

    \midrule
    300 & 50 & 328 & 0.911 & 0.65 & 0.883 & 0.74 \\
     &  & 794 & 0.796 & 0.71 & 0.771 & 0.88 \\
     &  & 1957 & 0.626 & 0.88 & 0.699 & 0.93 \\
     &  & 3550 & 0.089 & 0.97 & 0.287 & 0.96 \\
     &  & 4907 & 0.0 & - & 0.097 & 0.99 \\
     &  & 4912 & 0.0 & - & 0.097 & 0.99 \\
     &  & 5069 & 0.0 & - & 0.193 & 0.99 \\
     &  & 5106 & 0.0 & - & 0.030 & 0.74 \\
     &  & 5946 & 0.0 & - & 0.041 & 0.68 \\
     &  & 7040 & 0.016 & 0.75 & 0.0 & - \\
    \bottomrule
    \end{tabular}
\end{table*}

\section{Sensitivity to computation time of QKP instance}
\label{sec:appendixB}

\begin{figure}[t]
  \centering
  \begin{minipage}{0.46\linewidth}
    \includegraphics[width=\linewidth]{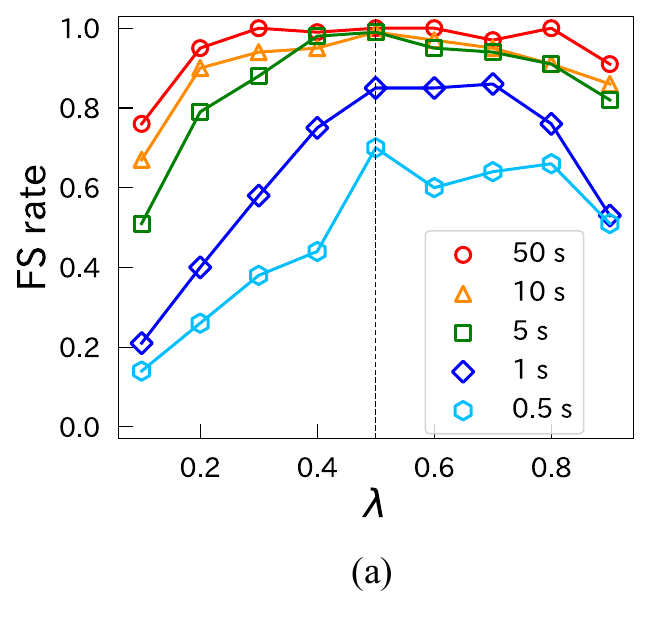}
  \end{minipage}
  \hspace{5mm}
  \begin{minipage}{0.46\linewidth}
    \includegraphics[width=\linewidth]{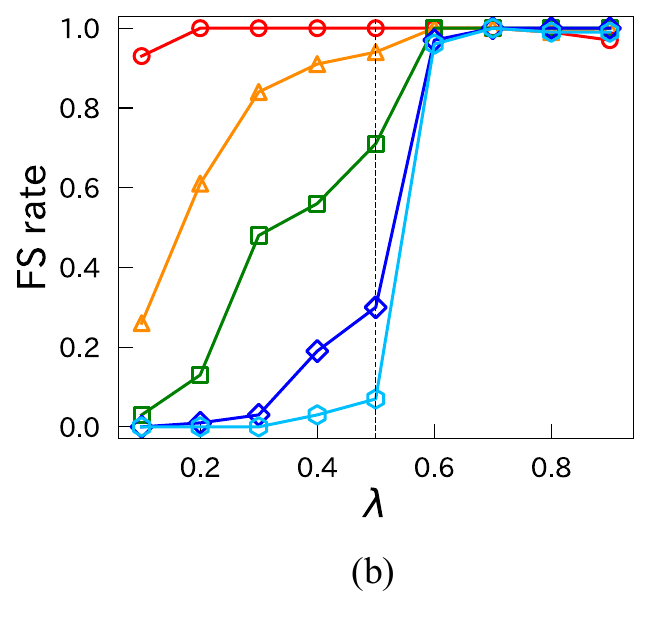}
  \end{minipage}
  \\
  \vspace{5mm}
  \begin{minipage}{0.46\linewidth}
    \includegraphics[width=\linewidth]{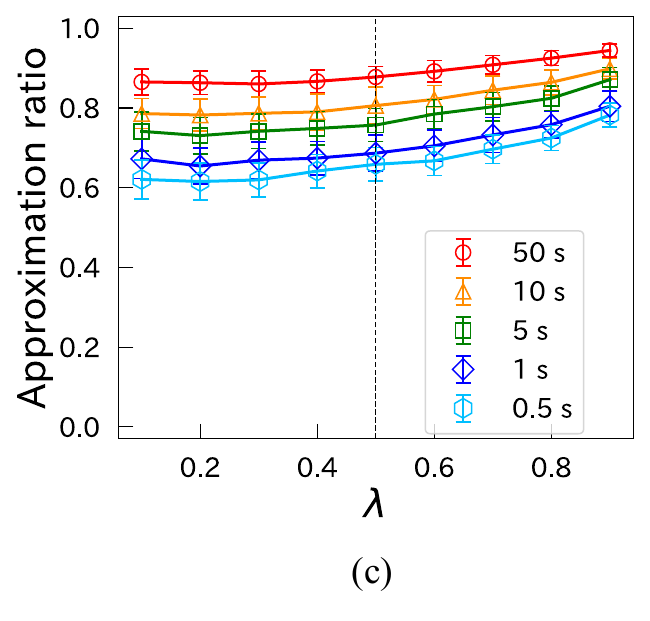}
  \end{minipage}
  \hspace{5mm}
  \begin{minipage}{0.46\linewidth}
    \includegraphics[width=\linewidth]{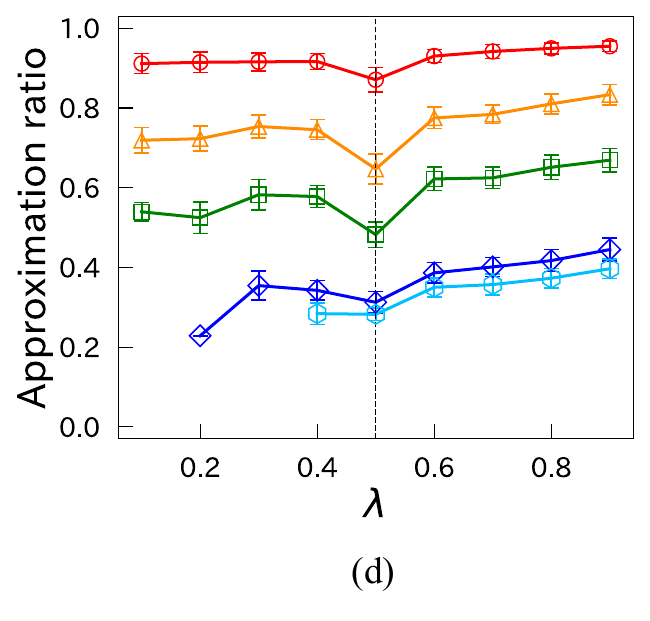}
  \end{minipage}
  \caption{Sensitivity to computation time of instance with knapsack capacity of $669$ under $ \left( n, d \right) = \left( 100, 25 \% \right)$. FS rates and approximation ratios are shown for one-hot encoding [(a), (c)] and domain-wall encoding [(b), (d)]. Sky blue hexagons, blue diamonds, green squares, orange triangles, and red circles represent computation times of $0.5, 1, 5, 10$, and $50$ s, respectively, for one-hot and domain-wall encoding. Dotted vertical lines denote $\lambda = 0.5$. Every plot of approximation ratio is average of $100$ runs. Error bars are standard deviations. Solid lines between points are merely a guide to the eye.}
  \label{fig:100_25_1_timeout}
\end{figure}

In this appendix, we assess the sensitivity of computation time for an instance with a knapsack capacity of $669$ under $ \left( n, d \right) = \left( 100, 25 \% \right)$.
Computation times were set to $0.5, 1, 5, 10$, and $50$ s.
The value of $\mu$ was fixed at $10$, and $\lambda$ was varied between $0.1$ and $0.9$ in increments of $0.1$, with $100$ runs performed for each setting.
Figure~\ref{fig:100_25_1_timeout} demonstrates that both one-hot and domain-wall encodings exhibit increases in FS rate and approximation ratio values as computation time increases.
Domain-wall encoding demonstrates a higher sensitivity to computation time, with more pronounced improvements as computation time extends.

\section*{Acknowledgment}
The authors would like to express their sincere gratitude to the World Premier International Research Center Initiative (WPI), MEXT, Japan, for their supporting of the Human Biology-Microbiome-Quantum Research Center (Bio2Q).

\bibliography{ref.bib}

\begin{IEEEbiography}[{\includegraphics[width=1in,height=1.25in,clip,keepaspectratio]{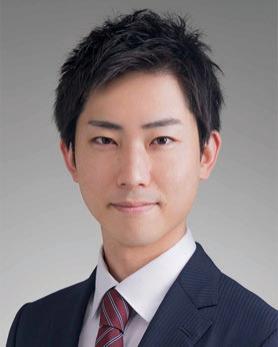}}]{Shuta Kikuchi} received the B.~Eng. and M.~Eng. degrees from the Waseda University in 2017 and 2019, and Dr.~Eng. degree from the Keio University in 2024, respectively. He is currently a Project Assistant Professor with the Graduate School of Science and Technology, Keio University. His research interests include Ising machine, statistical mechanics, and quantum annealing. He is a member of the JPS.
\end{IEEEbiography}

\begin{IEEEbiography}[{\includegraphics[width=1in,height=1.25in,clip,keepaspectratio]{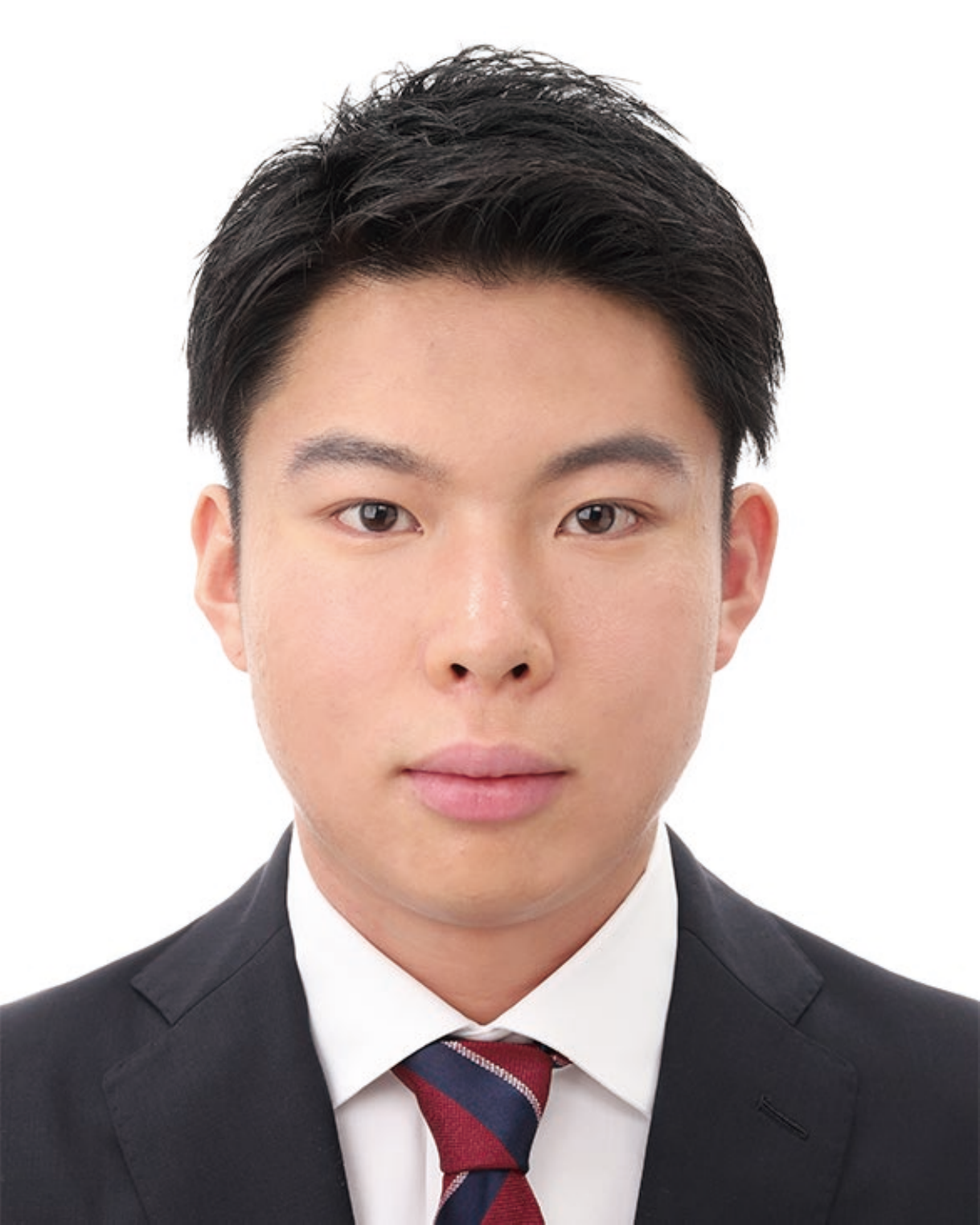}}]{Kotaro Takahashi} received the B.~Eng. and M.~Eng. degrees from the Keio University in 2022 and 2024, respectively. His research interests include quantum annealing and Ising machines.
\end{IEEEbiography}

\begin{IEEEbiography}[{\includegraphics[width=1in,height=1.25in,clip,keepaspectratio]{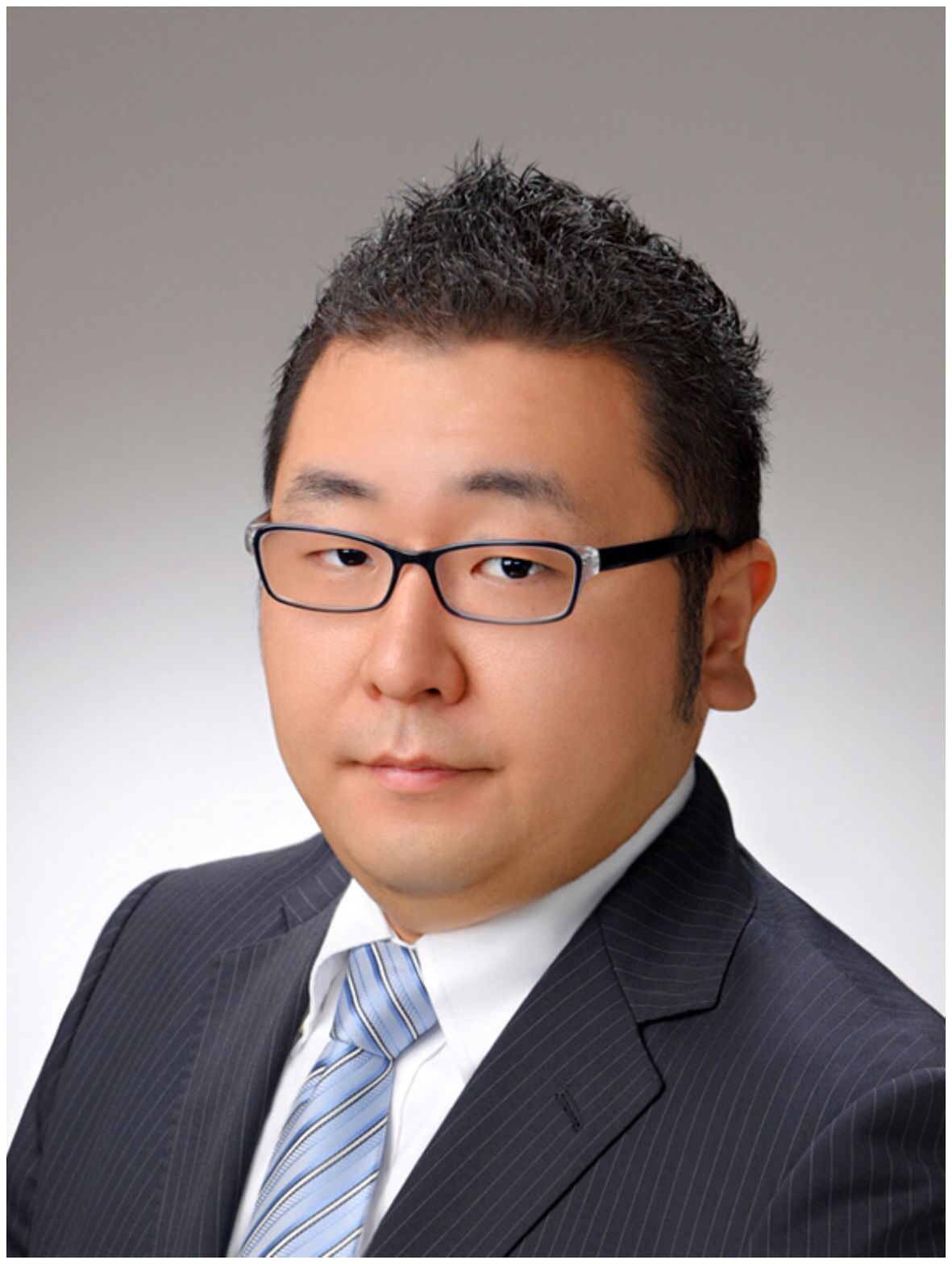}}]{Shu Tanaka} (Member, IEEE) received the B.~Sci. degree from the Tokyo Institute of Technology, Tokyo, Japan, in 2003, and the M.~Sci. and Dr.~Sci. degrees from The University of Tokyo, Tokyo, Japan, in 2005 and 2008, respectively. He is currently an Associate Professor with the Department of Applied Physics and Physico-Informatics, Keio University, a chair of the Keio University Sustainable Quantum Artificial Intelligence Center (KSQAIC), Keio University, and a Core Director at the Human Biology-Microbiome-Quantum Research Center (Bio2Q), Keio University. He is also a Visiting Associate Professor at the Green Computing Systems Research Organization (GCS), Waseda University. His research interests include quantum annealing, Ising machines, quantum computing, statistical mechanics, and materials science. He is a member of the Physical Society of Japan (JPS), and the Information Processing Society of Japan (IPSJ).
\end{IEEEbiography}

\EOD

\end{document}